\documentclass[conference]{IEEEtran}
\ifCLASSINFOpdf
\else
\fi

\usepackage{amsmath}
\usepackage{xcolor,colortbl}
\usepackage{graphicx}
\usepackage{dblfloatfix}
\usepackage{multirow}
\usepackage{multicol}
\usepackage{cite}
\usepackage{epstopdf}
\usepackage{subfigure}
\usepackage{url}
\usepackage{enumerate}
\usepackage{color}                    
\usepackage{amssymb}
\usepackage{amsmath}
\usepackage{xcolor,colortbl}
\usepackage{graphicx}
\usepackage{dblfloatfix}
\usepackage{multirow}
\usepackage{multicol}
\usepackage{cite}
\usepackage{epstopdf}
\usepackage{subfigure}
\usepackage{url}
\usepackage{enumerate}
\usepackage{color}                    
\usepackage{amssymb}

\newtheorem{remark}{Remark}

\title{ Cloud-Aided State Estimation of A Full-Car Semi-Active Suspension System}

        \author{\IEEEauthorblockN{Lixian Zhang$^1$, Xunyuan Yin$^2$, Junnan Shen$^1$, Haitao Yu$^3$}
\IEEEauthorblockA{$^1$Research Institute of Intelligent Control and System, Harbin Institute of Technology, Harbin, China 150080\\
$^2$Department of Chemical and Materials Engineering, University of Alberta, Edmonton, Canada T6G 2V2\\
$^3$State Key Laboratory of Robotics and Systems, Harbin Institute of Technology, China 150001\\
Email: lixianzhang@hit.edu.cn; xunyuan@ualberta.ca; jnshen@hit.edu.cn; yht@hit.edu.cn}}

\begin{document}

\maketitle

\begin{abstract}
In this work, we investigate a state estimation problem for a full-car semi-active suspension system. To account for the complex calculation and optimization problems, a vehicle-to-cloud-to-vehicle (V2C2V) scheme is utilized. Moving horizon estimation is introduced for the state estimation system design. All the optimization problems are solved in a remotely-embedded agent with high computational ability. Measurements and state estimates are transmitted between the vehicle and the remote agent via networked communication channels. The effectiveness of the proposed method is illustrated via a set of simulations.
\end{abstract}

\section{Introduction}
The automotive industry has witnessed a substantial increase in the popularity of advanced technologies, the implementations of which, however, are at the cost of extremely complex computing. While much effort has been devoted to the upgrading of on-board electronic control units (ECUs), a cloud computing agent provides a promising alternative \cite{Filev2013}. Cloud-aided computing has been widely adopted when addressing issues with respect to advanced automotive monitoring, control and optimization \cite{6818423,7287765,6974301,iet20162}, the realization of which is computationally demanding and requires extremely high computational efficiency. Based on a cloud-aided agent, a vehicle-to-cloud-to-vehicle (V2C2V) architecture has been constructed for the implementation of various automotive functions that can be hardly completed via ECUs. Within a V2C2V framework, a remote agent with high computational capability is exploited for complex calculations and optimizations. Networked communication links are used for data transmission between the vehicle and the remote agent. Please see \cite{Filev2013} for more detailed advantages.

Suspension and chassis systems constitute an important research topic in terms of automobiles as such systems substantially affect the safety, handling and riding comfort of a vehicle, which account for major concerns from both manufacturers and consumers. Due to the increasing demands in improved riding performance, great importance has been attached to the development of semi-active suspensions, which are able to provide a compromise in terms of handling, riding comfort and economic costs \cite{1709928,4729738}. To achieve improved riding performance, various control methodologies have been proposed, such as adaptive control \cite{974337}, fuzzy logic control \cite{995124}, $H_{\infty}$ control \cite{5291703,7452424}, etc. However, almost all existing designs were developed based on an assumption that all the system states are continuously available, which is challenging to realize from a practical perspective. In a V2C2V context, the computational complexity can be successfully handled as the embedded agent can be regarded as a source with enormous computing power.

In order to re-construct the unmeasured system states, state estimation should be carried out for the suspension control systems. In \cite{327121}, a Kalman filter was designed for the state estimation of a fully-active suspension. In \cite{Li2015275}, a filtering problem was investigated for suspension systems considering networked communication delays. However, it is worth mentioning that existing results only focus on a quarter-car suspension model, which cannot precisely reflect the dynamics of an entire vehicle. Moreover, these developed methods might not be directly applied to a full car suspension and chassis system. For example, the method proposed in \cite{Li2015275} could become infeasible due to the large scale of a full-car model.

For a class of linear systems, promising state estimation methodologies include Luenberger observer, Kalman filtering, moving horizon estimation (MHE), $H_{\infty}$ filtering \cite{Grewal2011,Liu2013376,7407325,7446321} , etc. Luenberger observer was exploited for deterministic systems, thus is not favourable for suspension systems subject to various disturbances \cite{1099826}. Kalman filter provides optimal estimates when the noise and disturbances are white \cite{kalman1960}, which is rarely satisfied in practice. Alternatively, MHE that gives state estimates by solving a batch least-square optimization problem is considered as a candidate for state estimation of suspension systems. MHE creates a moving window with a fixed size, within which the data are taken into account as available information \cite{5484538}. It has the capability to handle constraints and disturbances, which are always encountered in suspension and chassis systems. Based on above observations, MHE is adopted as the state estimation candidate in this work.

In this paper, we cope with a state estimation problem for a full-car chassis model with four suspensions.  A V2C2V framework is employed. An MHE-based state estimation algorithm is utilized to account for the estimation problem. The state estimator is assumed to be embedded in a remote agent. The vehicle is connected to the remote agent via wireless communication links as shown in Figure~\ref{network}. Specifically, the uneven road information (e.g., potholes) is stored in a remote agent as prior knowledge. The measurements collected by on-board sensors as well as real-time localization information of the vehicle are sent to the estimators in the remote agent. Based on the localization of the vehicle, the corresponding road profile information is retrieved and the state estimates are generated by solving the on-line optimization problem subject to certain constraints on the system states.

\textit{Notation:}\emph{\ } $\left\{w({t_p})\right\}^{k-1}_{p=0}$ is used to denote a sequence of $w$ with respect to $t$, i.e., $w({t_0}),~w({t_1}),~\ldots,w({t_k})$. The operator $\left|\cdot\right|_Q^2$ represents the square of the weighted Euclidean norm of a vector that is defined as $\left|x\right|_Q^2=x^TQx$. The remained notations used throughout this paper are fairly
standard.
\section{System Description and Problem Formulation}
In this section, we describe the cloud-aided vehicle state estimation and control framework, establish a model that depicts the dynamics of the full-car suspension system and formulate the state estimation problem.
\subsection{Cloud-aided state estimation and control framework}
The computational and data storage capability of the onboard computation units that serve as the key components of an electronic control unit (ECU) embedded in a vehicle is relatively limited. It is not recommended to carry out state estimation via the onboard computation units due to computational complexity concerns, especially when employing MHE-based algorithms for which optimization problems should be solved. To address this issue, we employ a vehicle-to-cloud-to-vehicle (V2C2V) framework realized via networked communications.

\begin{figure}[t]
\centering
 \includegraphics[width=0.5\textwidth]{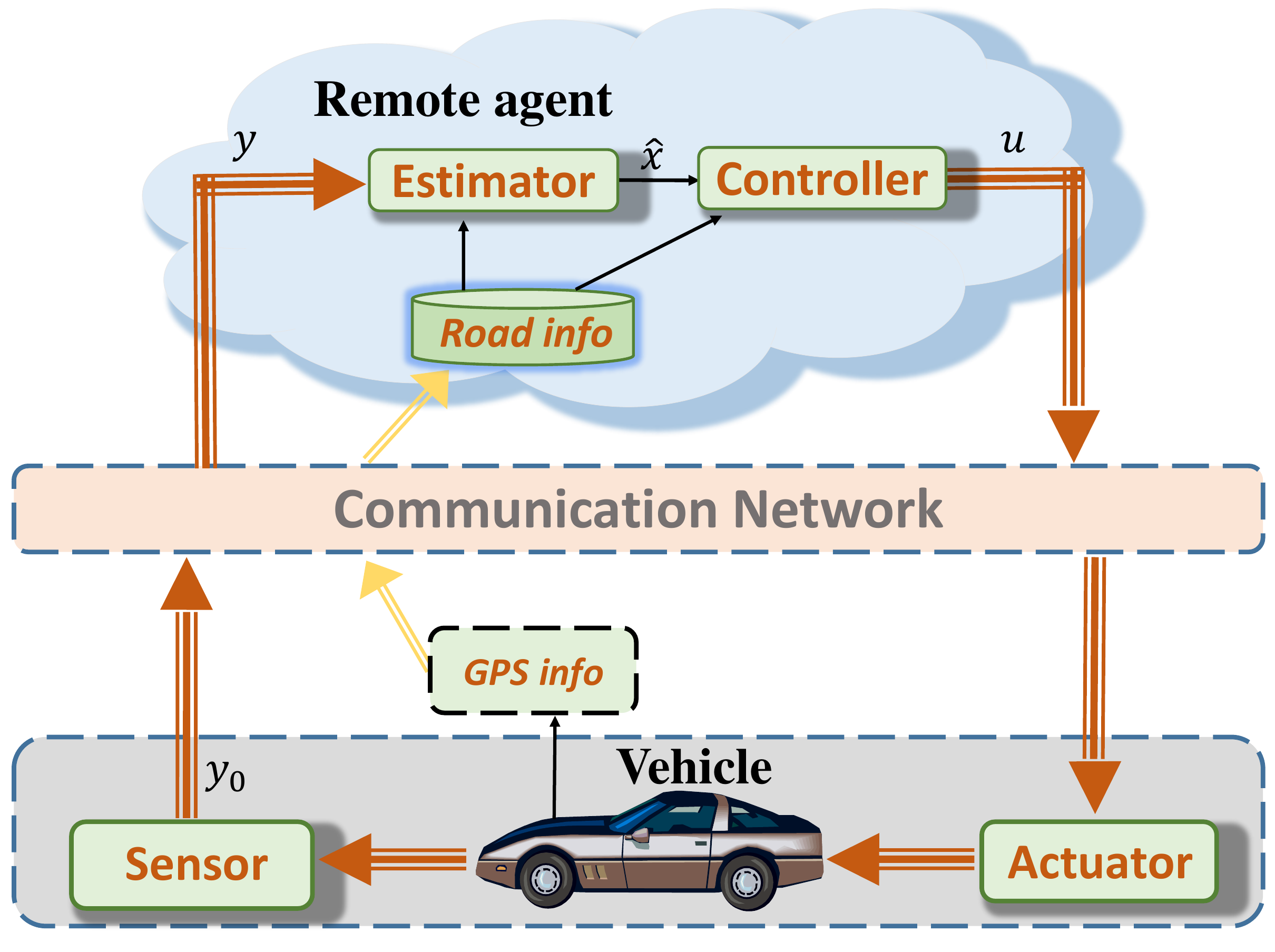}
      \caption{A diagram of the networked V2C2V framework.}
      \label{network}
\end{figure}
A V2C2V state estimation and control system is described as in Figure~\ref{network}. In a typical V2C2V system, a remote agent characterized by high computational, communicational and storage capabilities is exploited to account for the complex optimization computations and advanced automotive control. Therefore, the remote agent can be treated as a source of unlimited computational power and all necessary data. In the remote agent, road profiles are stored in advance. the information measured by the sensors in a vehicle is sent to this agent, where state estimates and control signals are generated by performing calculations and optimizations. We note the calculation requires the real-time matched road information that can be retrieved based on the GPS localization information received from the vehicle.

\begin{remark}
In this work, we only focus on how to accurately estimate the system states of the full-car suspension control system. Based on the obtained state estimates, it is not challenging to design a state estimation based suspension control method to more efficiently manipulate the suspensions for improved safety, handling and comfort.
\end{remark}
\subsection{Full-car suspension system}

\begin{figure}[t]
\centering
 \includegraphics[width=0.5\textwidth]{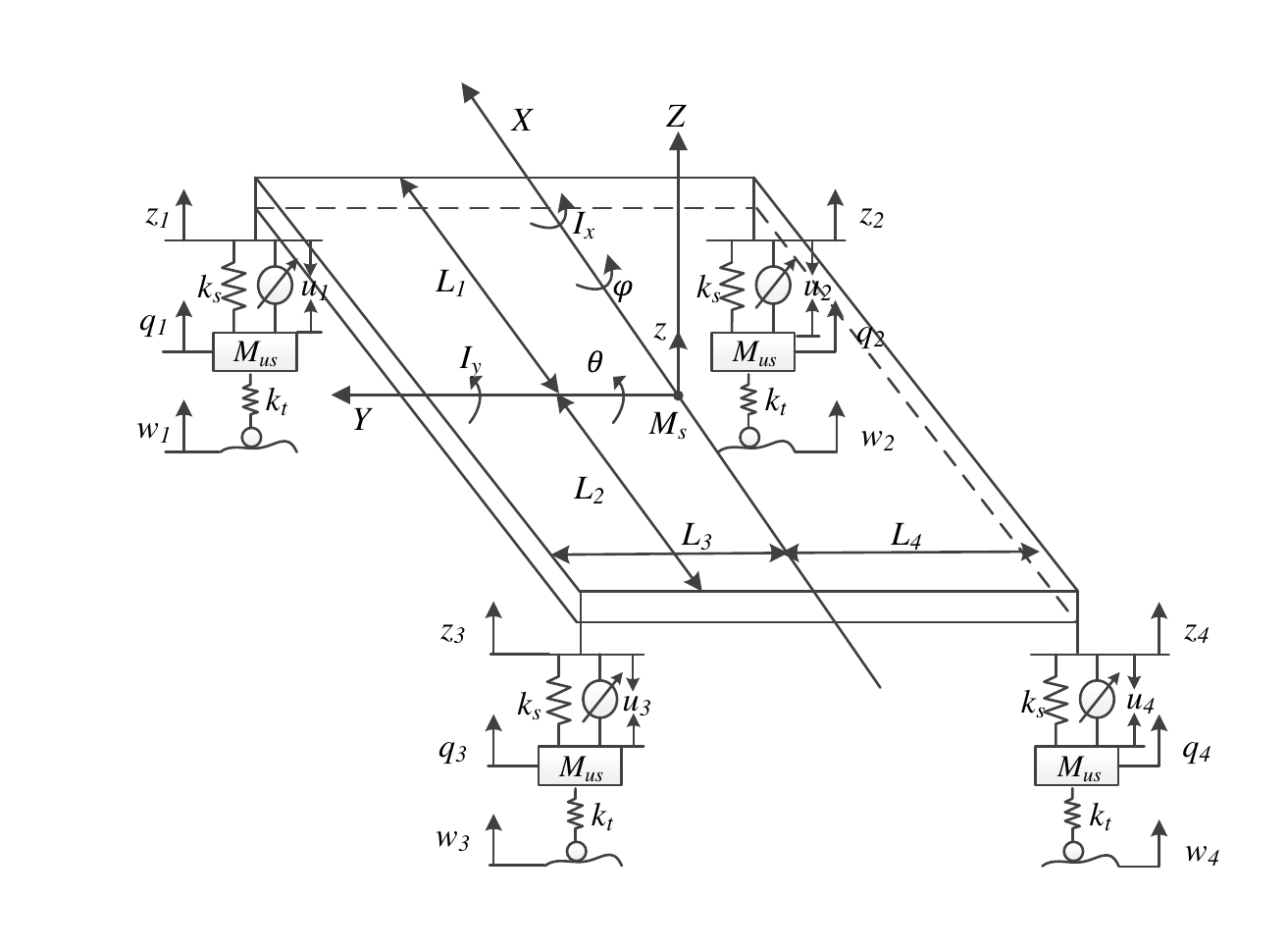}
      \caption{A schematic of the full-car suspension system.}
      \label{schematic}
\end{figure}

In this work, we takes into account a full-car semi-active suspension system with 7 degrees of freedom (DOF) \cite{Yagiz20081457}. The full-car model consists of four individual suspension systems, each of which is used to connect one wheel assembly to the vehicle body via a spring and a shock absorber with adjustable damping ratio. A schematic depicting the full-car suspension system is presented in Figure \ref{schematic}, while the definitions of the symbols used in the schematic are given in Table~\ref{tbl:variables}. Based on the schematic of the full-car system, the equations that describe the motions are obtained \cite{Yagiz20081457,7736990}:
\begin{enumerate}
\item{Front Left Wheel}
\begin{equation}\label{eq:1}
M_{us}\ddot{q}_1=-k_{t}(q_1-w_1)+F_1.
\end{equation}
\item{Front Right Wheel}
\begin{equation}\label{fr}
M_{us}\ddot{q}_2=-k_{t}(q_2-w_2)+F_2.
\end{equation}

\item{Rear Left Wheel}
\begin{equation}\label{rl}
M_{us}\ddot{q}_3=-k_{t}(q_3-w_3)+F_3.
\end{equation}

\item{Rear Right Wheel}
\begin{equation}\label{rr}
M_{us}\ddot{q}_4=-k_{t}(q_4-w_4)+F_4.
\end{equation}

\item{CG-vertical}
\begin{equation}\label{cg}
M_s\ddot{z}=-F_1-F_2-F_3-F_4.
\end{equation}

\item{Pitch}
\begin{equation}
I_y\ddot{\theta}=L_1F_1+L_1F_2-L_2F_3-L_2F_4.
\end{equation}

\item{Roll}
\begin{equation}\label{eq:7}
I_x\ddot{\phi}=-L_3F_1+L_4F_2-L_3F_3+L_4F_4.
\end{equation}
\end{enumerate}
where
\begin{equation*}
\begin{aligned}
z_1=&z-L_1\theta+L_3\phi, ~~z_2=z-L_1\theta-L_4\phi,\\[0.26em]
z_3=&z+L_2\theta+L_3\phi, ~~z_4=z+L_2\theta-L_4\phi,\\[0.26em]
F_1=&k_s(z_1-q_1)+c_s(\dot z_1-\dot q_1)+u_1,\\[0.26em]
 F_2=&k_s(z_2-q_2)+c_s(\dot z_2-\dot q_2)+u_2,\\[0.26em]
F_3=&k_s(z_3-q_3)+c_s(\dot z_3-\dot q_3)+u_3,\\[0.26em]
F_4=&k_s(z_4-q_4)+c_s(\dot z_4-\dot q_4)+u_4
\end{aligned}
\end{equation*}

To construct a system model of differential equations for the full-car suspension control system, we select 14 system states which we are concerned with to construct the state vector, i.e., $x=\left[x_1~x_2~\ldots~x_{14}\right]^T$ with $x_1=q_1$, $x_2=\dot q_1$, $x_3=q_2$, $x_4=\dot q_2$, $x_5=q_3$, $x_6=\dot q_3$, $x_7=q_4$, $x_8=\dot q_4$, $x_9=z$, $x_{10}=\dot z$, $x_{11}=\theta$, $x_{12}=\dot {\theta}$, $x_{13}=\phi$ and $x_{14}=\dot{\phi}$.
We consider that the disturbances to the suspension system mainly consist of two parts: the uneven road conditions (e.g., potholes) and unknown disturbances (mainly GPS localization errors). Defining $\dot r$ as the the uneven road information which is known as prior knowledge stored in the remote agent and define $\bar w$ as the unknown disturbances, the overall disturbances can be considered separately, i.e, $w=\dot r +\bar w$.
The system model is then established as
\begin{equation}\label{eq:model}
\dot x=Ax+Bu+B_r{\dot r}+B_r\bar w
\end{equation}
where $\dot r=\big[\dot r_1~\dot r_2~\dot r_3 ~\dot r_4\big]^T$ depicts the affects on the dynamics of the four suspensions by time-varying road conditions,
 $\bar w=\big[\bar w_1~\bar w_2~\bar w_3 ~\bar w_4\big]^T$ contains unknown disturbances to the four suspensions, $A$, $B$ and $B_r$ are constant system matrices and can be calculated via Eq. (\ref{eq:1})-(\ref{eq:7}), thus are omitted for brevity.
\subsection{Formulation of the state estimation problem}
It is desirable to know the real-time information with respect to all the system states. Considering the vehicles equipped with semi-active suspension systems, it is easy to measure the sprung/unsprung mass velocities, pitch/roll angle rate of the car body. However, the rest system states are expensive to measure and cannot be directly obtained by conducting integration on the measurable states and the errors inevitably accumulate with respect to time.

Based on above observations, we assume that the output vector of system (\ref{eq:model}), without considering measurement noises, is
\begin{equation}\label{eq:output}
\begin{aligned}
y_0&={\left[ {\begin{array}{*{10}{c}}
{{x_2}}&{{x_4}}&{{x_6}}&{{x_8}}&{{x_{10}}}&{{x_{12}}}&{{x_{14}}}
\end{array}} \right]^T}\\[0.2em]
\end{aligned}
\end{equation}
From (\ref{eq:model}) and (\ref{eq:output}), we have
\begin{equation}
\begin{aligned}
\dot x(t)&=Ax(t)+Bu(t)+B_r{\dot r(t)}+B_r\bar w(t)\\
y_0&=C x(t)+ D v(t)
\end{aligned}
\end{equation}
where $C$ is a constant matrix that can be determined following (\ref{eq:output}), $D$ is set to be an identity matrix and $v$ denotes a vector of measurement noises.

The objective is to estimate all the unmeasurable system states based on the available system measurements with existence of external disturbances induced by uneven road conditions and inaccurate GPS localization.

\begin{table}[t]
\caption{Suspension system variables}
\renewcommand\arraystretch{1.35}
\label{tab1}\centering%
\begin{tabular}{lp{0.55\columnwidth}}
  \hline\noalign{\smallskip}
   $q_{i},~i=1,~2,~3,~4$ & unsprung mass displacements \\
   $z_{i},~i=1,~2,~3,~4$ & sprung mass displacements \\
   $w_{i},~i=1,~2,~3,~4$ & external disturbances \\
   $u_{i},~i=1,~2,~3,~4$ & adjustable  damper forces \\
   $z$ & displacement at center of gravity\\
   $M_s$ & car body sprung mass\\
   $M_{us}$ & tire and axles unsprung mass\\
   $k_{s}$ & suspension stiffness\\
   $k_{t}$ & tire stiffness\\
   $c_{s}$ & damping constant\\
   $\phi$ & roll angle of the car body\\
   $\theta$ & pitch angle of the car body\\
   $I_x$ & mass moment of inertia for roll motions\\
   $I_y$ & mass moment of inertia for pitch motions\\
   $L_1$ & vertical distance between front suspensions and center of gravity\\
   $L_2$ & vertical distance between rear suspensions and center of gravity\\
   $L_3$ & horizonal distance between left suspensions and center of gravity\\
   $L_4$ & horizonal distance between right suspensions and center of gravity\\
  \noalign{\smallskip}\hline
\end{tabular}\label{tbl:variables}
\end{table}

\section{MHE-Based State Estimation Design}
Constraints on system states, manipulated inputs and external disturbances are frequently encountered in suspension systems. MHE has inherent advantages in terms of handling explicit constraints compared with other state estimation methodologies. In this work, an MHE-based estimator is utilized to approximate the system states of the full-car suspension system.

The MHE algorithm gives state estimates by solving an optimization problem in a discrete-time context.
\subsection{Full-information state estimation}
We first present some background knowledge that is essential for the development of MHE algorithms. Suppose that we consider all the available measurements at previous sampling instants, a full-information state estimation scheme is formulated, of which the corresponding objective function can be defined as \cite{Rao20011619}:
\begin{equation}\label{objective1}
\begin{aligned}
{\Theta _{{t_k}}}&\left({ x({t_0}),\left\{{ \bar w({t_p})}\right\}^{k-1}_{p=0}}\right)\\[0.23em]
= &\sum\limits_{p = 0}^{k - 1} {\left|\bar w(t_p)\right|^2_{Q^{-1}} +\sum\limits_{p = 0}^{k } \left| v(t_p)\right|^2_{R^{-1}} } \\[0.3em]
&  + {\left(x({t_0}) - \hat x({t_0})\right)^T}\Pi _0^{ - 1}\left(x({t_0}) - \hat x({t_0})\right)
\end{aligned}
\end{equation}
with
$$v(t_p)\triangleq y(t_p)-Cx(t_p),~p=0,~1,~\ldots,~k$$
In (\ref{objective1}), $Q$ and $R$ and symmetric positive definite matrices with compatible dimensions, $\hat x$ denotes a state estimate of $x$, $\left(\hat x_0,~\Pi_0\right)$ is a pair summarizing the prior information at initial time instant $t_0$. The state estimation problem subject to constraints can be converted to a minimization problem with respect to objective function (\ref{objective1}) as follows:
\begin{equation}\label{eq:optimal}
{\Theta^* _{{t_k}}} =\mathop {\min }\limits_{x({t_0}),\left\{{\bar w({t_p})}\right\}^{k-1}_{p=0}} {\Theta _{{t_k}}}\left({ x({t_0}),\left\{{\bar w({t_p})}\right\}^{k-1}_{p=0}}\right)
\end{equation}
which is subject to constraints
\begin{equation*}
x(t_p)\in\mathbb X,~~\bar w(t_p)\in \mathbb W,~~v(t_p)\in \mathbb V,~~p=0,~1,\ldots,~k
\end{equation*}
where symbols $\mathbb X$, $\mathbb W$ and $\mathbb V$ denote polyhedral and convex sets that bound the states, disturbances and measurement noises, respectively.

By solving optimization problem (\ref{eq:optimal}), one is able to find a unique solution to $\hat x(t_0),~\hat{\bar w}(t_0),~\hat{\bar w}(t_1),~\ldots,~\hat{\bar w}(t_{k-1})$, based on which we can obtain the optimal sequence of state estimates
$$\hat x_{p|t_k}\triangleq x(p,\hat x_{t_0|t_k},\left\{\hat {\bar w}_p\right\}),~p=0,~1,~\ldots,~k-1$$

We note that with the increase in time, the computational burden will increase dramatically, which renders this algorithm infeasible for practical issues.
\subsection{Moving horizon estimation (MHE)}
To make the aforementioned algorithm applicable for state estimation purposes, the problem size should be restricted. MHE has been developed by reducing the full-information minimization problem (\ref{eq:optimal}) to a finite horizon quadratic problem \cite{Alessandri20081753,Rao20011619,1178905}. MHE explicitly takes into account the data within a horizon with a fixed dimension, while approximate all previous information (priori to the estimation horizon) that is not explicitly considered in the state estimator.

Based on this consideration, the objective function in (\ref{objective1}) can be split into two sections according to different time intervals, i.e., ${t_0} \le t < {t_{k - N}}$ and ${t_{k - N}} \le t \le {t_k}$. Specifically, objective function (\ref{objective1}) is re-considered as follows:
\begin{equation}\label{objective2}
\begin{aligned}
{\Theta _{{t_k}}}&\left({ x({t_0}),\left\{{\bar w({t_p})}\right\}^{k-1}_{p=0}}\right)\\[0.23em]
= &\sum\limits_{p = k-N}^{k - 1} { { \left|\bar w(t_p)\right|^2_{Q^{-1}} +\sum\limits_{p = k-N}^{k }  \left|v(t_p)\right|^2_{R^{-1}}} }\\[0.3em]
&  +\Theta _{t_{k-N}}\left({ x({t_0}),\left\{{ \bar w({t_p})}\right\}^{k-N}_{p=0}}\right)
\end{aligned}
\end{equation}

Next, we define a function
\begin{equation*}
V({t_{k - N}}) \triangleq \min {\Theta _{{t_{k - N}}}}\left( {x({t_0}),\left\{ {\bar w({t_p})} \right\}_{p = 0}^{k - N}} \right)
\end{equation*}
as the arrival cost of the MHE problem, which is also subject to the constraints on systems states and external disturbances.
Then, the moving horizon optimization problem for the suspension system is formulated as follows:
\renewcommand{\arraystretch}{1.36}
\renewcommand{\arraycolsep}{3pt}
\begin{subequations}\label{mhefast}
\begin{align}
&\mathop {\min }\limits_{\hat x({t_0}),\left\{{ \bar w({t_p})}\right\}^{k-1}_{p=k-N}} \left\{ \begin{array}{l}
\sum\limits_{p = k-N}^{k - 1}  \left|\bar w(t_p)\right|^2_{Q^{-1}}\\[0.3em]
~~~+\sum\limits_{p = k-N}^{k } \left|v(t_p)\right|^2_{R^{-1}}\\[0.9em]
 ~~~+V(t_{k-N})
\end{array} \right\}\label{mhefast:cost}\\[0.3em]
&\textmd{ s.t.~}\quad\dot {\hat x}(t)=A\hat x(t)+Bu(t_p)+B_r{\dot r(t_p)}+B_r\bar w(t_p),\notag\\[0.3em]
&~~~~~~~~~~~~~~~~t\in[t_p,t_{p+1}],\;p=k-N,\ldots,k-1,\label{mhefast:model}\\[0.3em]
& v(t_p)\triangleq y(t_p)-C\hat x(t_p),~p=k-N,~1,~\ldots,~k,\label{mhefast:v} \\[0.3em]
&\hat x(t_p)\in\mathbb X,~\bar w(t_p)\in \mathbb W,~v(t_p)\in \mathbb V,~~p=k-N,\ldots,~k. \label{mhefast:cons}
\end{align}
\end{subequations}
where $N$ denotes the fixed size of the estimation horizon, $Q$ and $R$ are symmetric positive definite matrices depicting the covariances of $\bar w$ and $v$, respectively, $V(t_{k-N})$ is the defined arrival cost summarizing all previous information up to time instant $t_{k-N}$ priori to the estimation horizon. In optimization problem (\ref{mhefast}), (\ref{mhefast:cost}) is the objective function to be minimized, (\ref{mhefast:model}) depicts the suspension system model,
(\ref{mhefast:cons}) describe the constraints on system states, disturbances as well as measurement noises.
\begin{remark}
It is generally challenging to derive exact mathematical expressions for the arrival cost when constraints exist. To cope with the arrival cost in a constrained extent, one practical solution is to approximate arrival cost $V(t_{k-N})$ based the Kalman filter following \cite{Rao20011619}.
\end{remark}
\section{Simulation Results}
In this section, we carry out a set of simulations to show state estimation results for the full-car suspension system.

The parameters of the variables of suspension system model (\ref{eq:1})-(\ref{eq:7}) are given in Table~\ref{tb:constants}.
\begin{table}[t]
\caption{Parameters of suspension system variables}
\renewcommand\arraystretch{1.35}
\label{tab1}\centering%
\begin{tabular}{lp{0.36\columnwidth}}
  \hline\noalign{\smallskip}
    ~~~~~~$I_x=4000\;Kg\cdot m^2$  & ~~$I_y=950\;Kg\cdot m^2$  \\
  ~~~~~ $L_1=1.4\;m$  & ~~$L_2=1.6~m$ \\
   ~~~~~~$L_3=1.0\;m$  & ~~$L_4=1.0~m$ \\
   ~~~~~~$M_s=1200\;Kg$ & ~~$M_{us}=60~Kg$ \\
  ~~~~~ ~$k_s=16800~N/m$ & ~~$k_t=190000~N/m$ \\
  ~~~~~ ~$c_s=800~N\cdot s/m$ \\
  \noalign{\smallskip}\hline
\end{tabular}\label{tb:constants}
\end{table}
\begin{figure}[t]
\centering
 \includegraphics[width=0.5\textwidth]{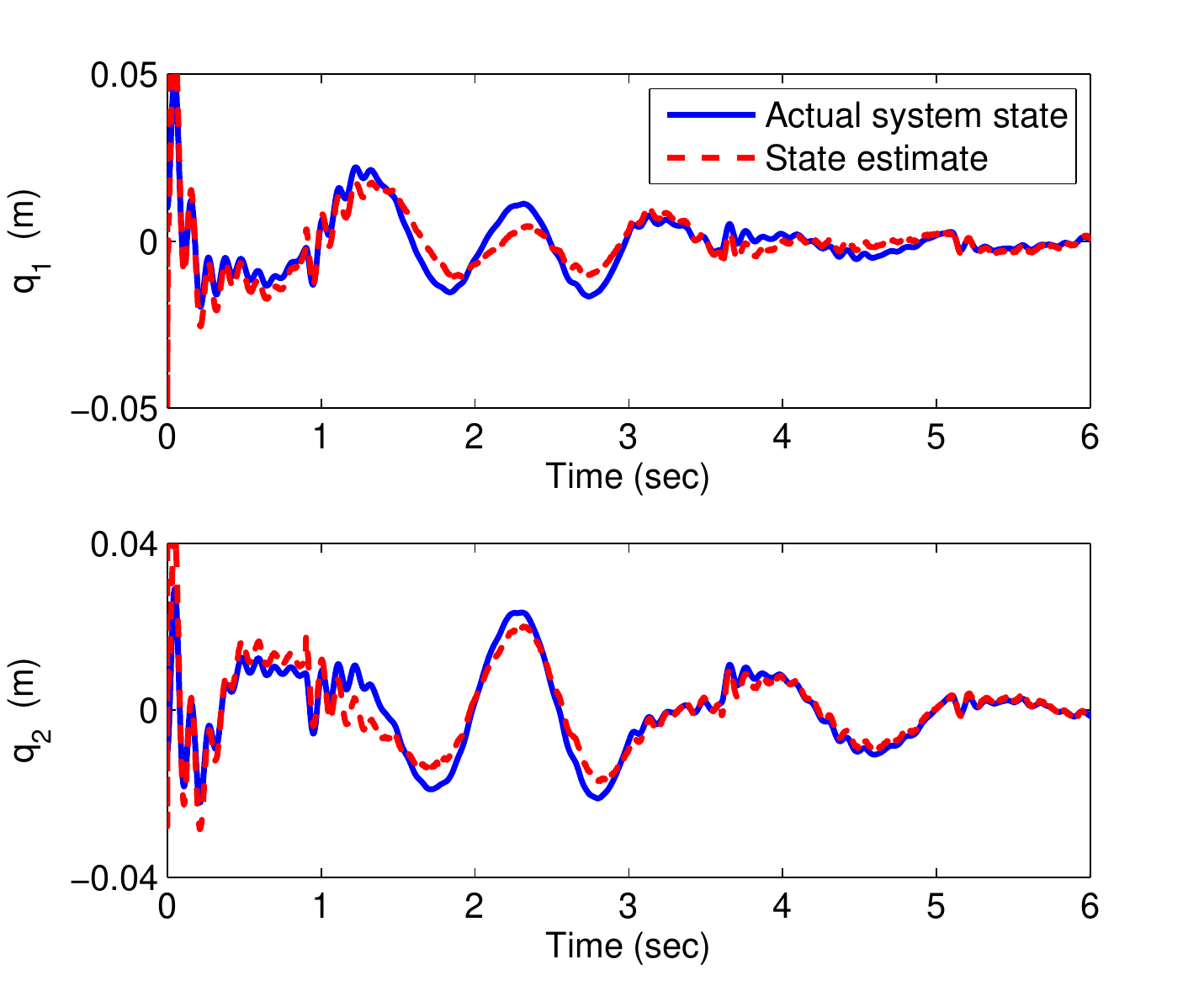}
      \caption{State estimates and actual states of unsprung mass displacements of suspension 1 and suspension 2.}
      \label{suspension1}
\end{figure}
\begin{figure}[t]
\centering
 \includegraphics[width=0.5\textwidth]{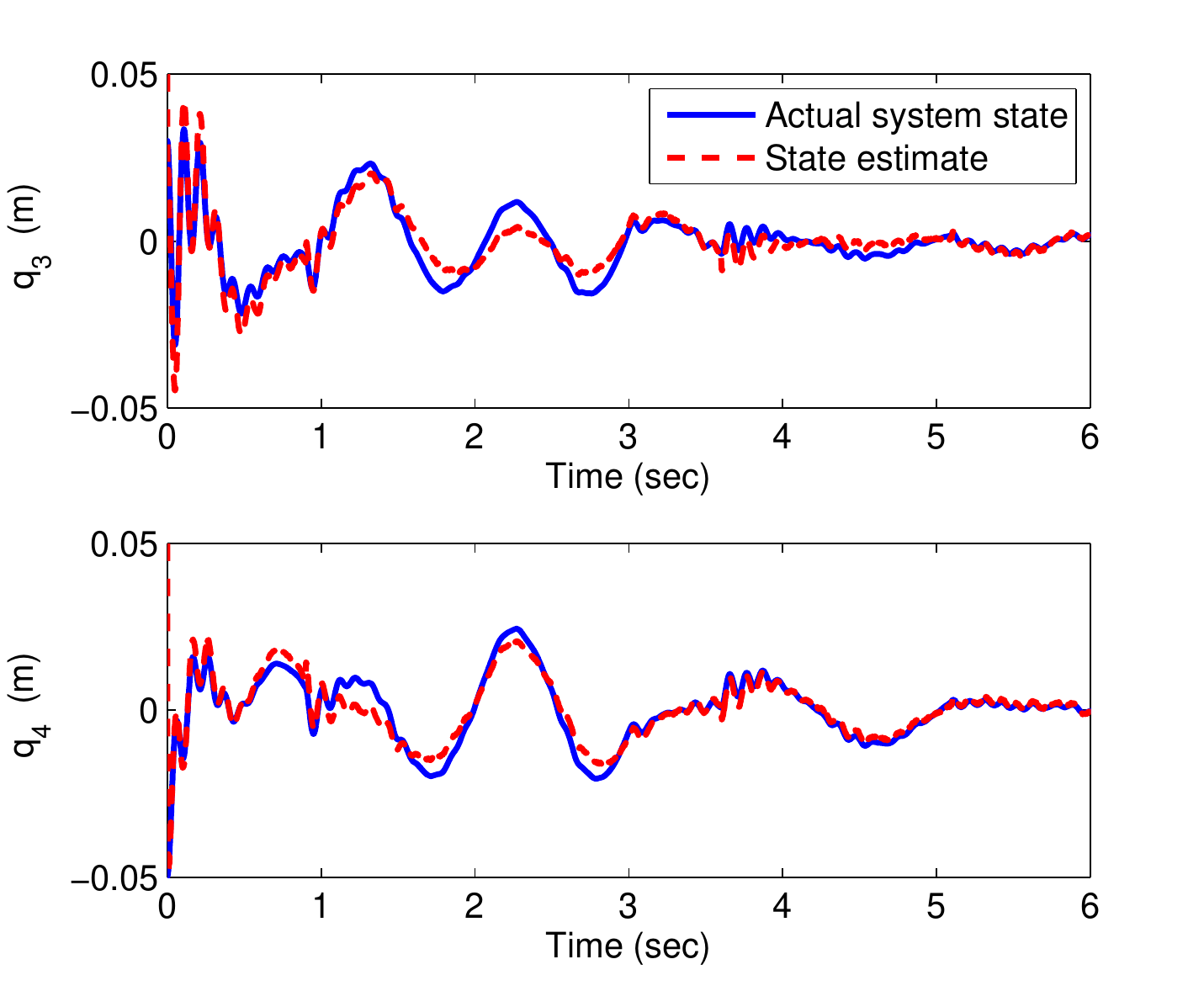}
      \caption{State estimates and actual states of unsprung mass displacements of suspension 3 and suspension 4.}
      \label{suspension2}
\end{figure}
\begin{figure}[t]
\centering
 \includegraphics[width=0.5\textwidth]{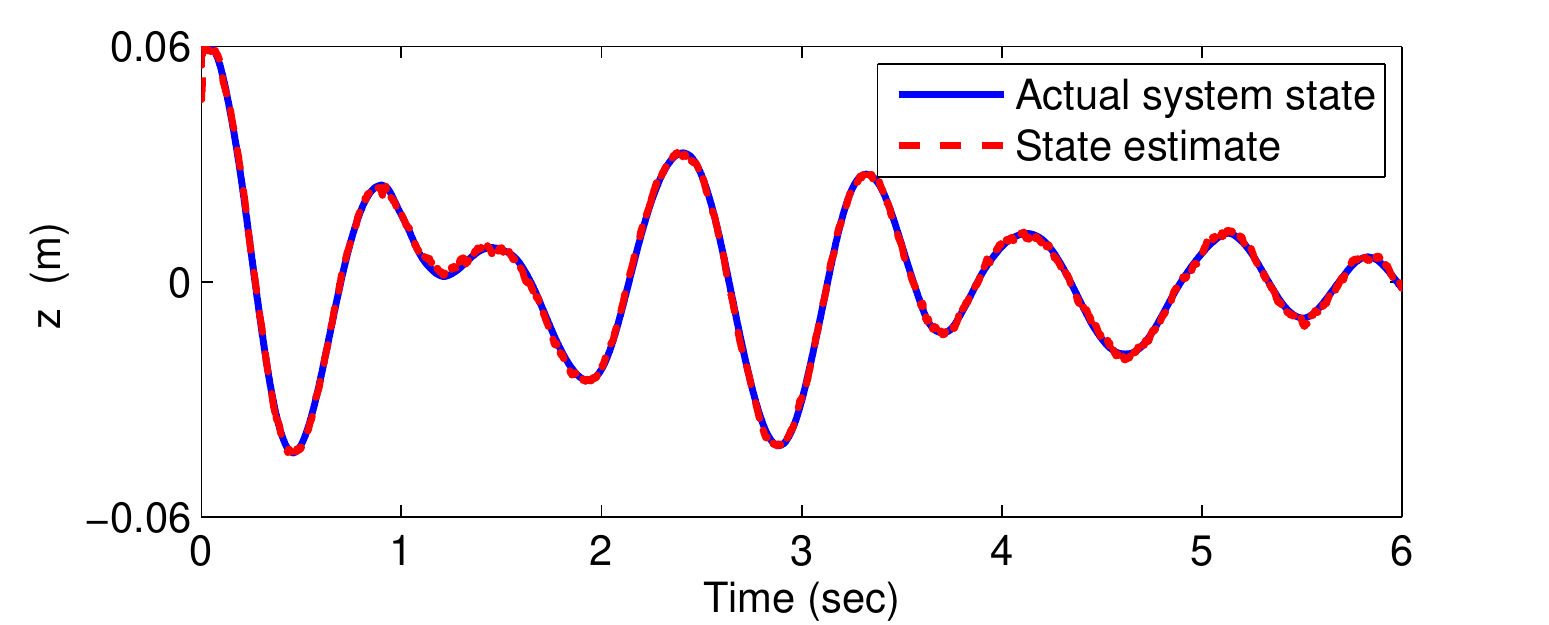}
      \caption{State estimates and actual states of the displacement at center of gravity.}
      \label{suspension3}
\end{figure}
\begin{figure}[t]
\centering
 \includegraphics[width=0.5\textwidth]{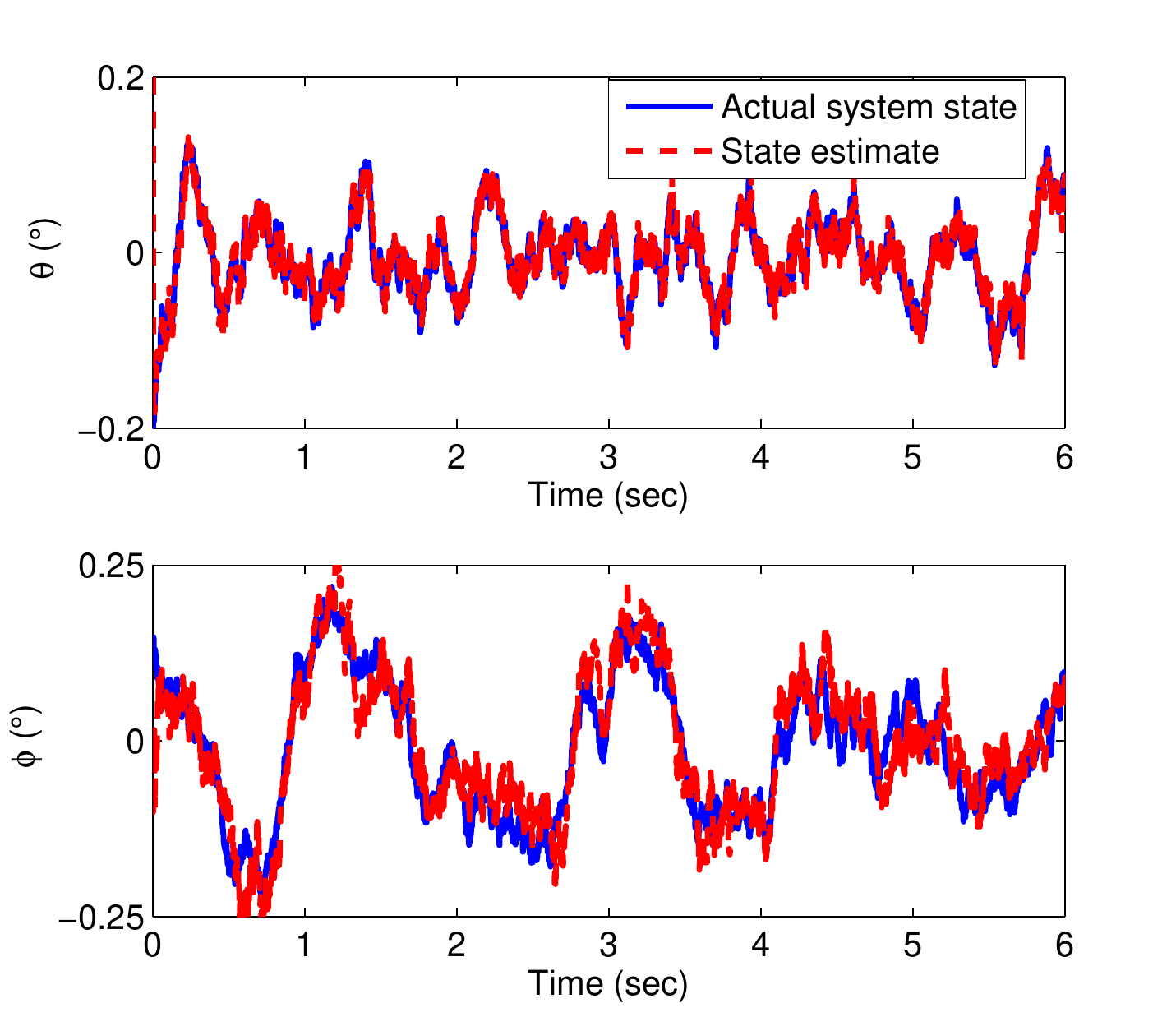}
      \caption{State estimates and actual states with respect to the vehicle attitude.}
      \label{suspension4}
\end{figure}
We carry out state estimation over a period of 6 sec. For simulation purposes, the uneven road information is modelled as the following segment:
\begin{equation*}
\dot r(t) = \left\{ \begin{array}{l}
2.58\times 10^{-2} \cdot \sin 2\pi t~~~~~~0.9s \le t \le 3.0s\\
1.23\times 10^{-2} \cdot \sin 1.2\pi t~~~~3.6s \le t \le 5.1s\\
0~~~~~~~~~~~~~~~~~~~~~~~~~~~~~~\text{otherwise}
\end{array} \right.
\end{equation*}
The uncertain system disturbances to all the four individual suspensions are assumed to be white noises.

In the optimization problem (\ref{mhefast:cost}), the weighing matrices are selected as $Q=diag\left\{0.25~1~0.25~1~0.25~1~0.25~1~0.3~1~0.5~0.5~0.5~0.5\right\}$ and $R=diag\left\{0.75~ 0.75 ~0.75 ~ 0.75~ 1~ 1 ~1\right\}$, respectively. The initial states of the suspension system are set to be $x_1(t_0)=0.01~m$, $x_2(t_0)=-0.1~m/s$, $x_3(t_0)=-0.01~m$, $x_4(t_0)=0.1~m/s$, $x_5(t_0)=0.03~m$, $x_6(t_0)=0.2~m/s$,$x_7(t_0)=-0.08~m$, $x_8(t_0)=0.2~m/s$, $x_9(t_0)=0.06~m$, $x_{10}(t_0)=0.04~m/s$, $x_{11}=-5 ^\circ$, $x_{12}=2^\circ /s$,$x_{13}=2 ^\circ$, $x_{14}=-3^\circ /s$.

Applying the state estimation algorithm in (\ref{mhefast}), the simulation results are obtained and given from Figure~\ref{suspension1} to Figure~\ref{suspension4}. For brevity, only the state estimates for the unmeasurable states are given. From the results we see that the state estimates in terms of both vertical displacements and the body attitude given by the developed scheme are able to accurately track actual system states, which verifies the effectiveness and applicability of this method.

From Figure~\ref{suspension1} and Figure~\ref{suspension2}, it is seen that the vertical displacements of the four individual suspensions experience similar trends, this is mainly because the same road profile information is added for the four suspensions in this set of simulations for the sake of effectiveness verification. We note that the road profile information for different suspensions could be different in practice, which depends on the accurate position of the vehicles on a specific road.
\begin{remark}
In Figure \ref{suspension3}, we see that the trajectory is smoother than those of other system states, which show that the disturbances to the individual suspensions to do not necessarily affect the overall vertical displacement of the chassis directly. This important property will also be utilized for the output-feedback control design in future works.
\end{remark}
\begin{remark}
In this framework, we assume the communication links are in ideal condition. In practice, however, time-varying delays are frequently encountered in the wireless data transmission phase. As long as the delays are sufficiently small, then the estimation performance can still be acceptable according to an additional set of simulations. If the time delays are large, then we may exploit a state predictor as designed in \cite{Zhang2014672} when measurements are unavailable, which constitutes a part of our future works.
\end{remark}

\section{Concluding Remarks}
In this study, we addressed a state estimation problem for a full-car chassis model with four individual semi-active suspensions. A full-car chassis and suspension system model was established. An MHE-based algorithm was used to design the state estimator. A V2C2V structure was constructed such that all the complex calculations and optimizations are carried out in a remote agent instead of the on-board ECUs. Due to the unlimited computing power of the remote agent, the calculational complexity of the optimization programs has been successfully handled. The effectiveness of the designed state estimation system with in the V2C2V framework has been verified. To improve the applicability of this networked estimation scheme, we will take into account the communicational imperfections including time delays, packet dropouts that may deteriorate the estimation performance in future works. We will also investigate an estimation-based output feedback control system realized via V2C2V as shown in Figure~\ref{network} to facilitate its applications.
\bibliographystyle{ieeetr}

\end{document}